\documentclass[aps,prl,showpacs,superscriptaddress,nofootinbib]{revtex4}
\usepackage{bbm}
\usepackage{mathrsfs}
\usepackage{epsfig,psfrag}
\usepackage{amsmath,amsfonts,amssymb}
\usepackage[usenames]{color}
\usepackage{bm}
\usepackage{bbm}

\usepackage{ulem}
\newcommand{\tr}{{\rm tr}}
\newcommand{\csch}{{\rm \, csch}}

\definecolor{mfm}{rgb}{.8,.08,.05}

\definecolor{ng}{rgb}{.6,.18,.15}

\begin{document}
\title{Electromagnetic Casimir Energies of Semi-Infinite Planes}


\author{Mohammad F. Maghrebi}
\email{magrebi@mit.edu}
\affiliation{Department of Physics,
Center for Theoretical Physics,
Massachusetts Institute of
Technology, Cambridge, MA 02139, USA}

\author{Noah Graham}
\email{ngraham@middlebury.edu}
\affiliation{Department of Physics,
Middlebury College,
Middlebury, VT 05753, USA}

\begin{abstract}
Using recently developed techniques based on scattering theory, we
find the electromagnetic Casimir energy for geometries involving
semi-infinite planes, a case that is of particular interest in the
design of microelectromechanical devices.  We obtain both approximate
analytic formulae and exact results requiring only modest numerical
computation.  Using these results, we analyze the effects of edges and
orientation on the Casimir energy.  We also demonstrate the accuracy,
simplicity, and utility of our approximation scheme, which is based on
a multiple reflection expansion.
\end{abstract}

\maketitle

\paragraph{Introduction and Method}

The Casimir energy has been most commonly studied using techniques
based on the original calculation for infinite parallel
planes \cite{Casimir48}. Recently developed techniques
\cite{spheres,universal} have made it possible to calculate the
electromagnetic Casimir force between objects of arbitrary shape and
electromagnetic response. In recent application{s} of these methods,
the interaction energy of a semi-infinite plane with an infinite
plane was analyzed in two different ways: First, with
the half-plane considered as the limit of a parabolic cylinder of zero
curvature \cite{parabolic}, in which case the exact energy can be
computed numerically. Second, with the semi-infinite plane taken as
the limit of a wedge of zero opening angle \cite{wedge}, in which case
it is convenient to consider a multiple reflection expansion for the
energy.  Remarkably, the analytic
formulae obtained by keeping the lowest few orders in the reflection
expansion give very good agreement with the full numerical result.
(This approach can also be extended to interactions of more than two
bodies \cite{Maghrebi10-2}.)  We study the Casimir interaction of
semi-infinite planes by contrasting these two methods.  This problem
is applicable to the design of microelectromechanical devices
\cite{Khoshnoud} and has been of recent theoretical interest as well
\cite{Gies,Kabat}.  We present analytic formulae obtained through the
multiple-reflection approximation and use the exact numerical
calculation to obtain a concrete measure of the accuracy of the
approximations involved.  This system provides an ideal environment in
which to study the effects of edges and orientation.

In the scattering theory approach \cite{spheres,universal}, the
Casimir interaction energy is expressed in terms of the scattering
$T$-matrices, also known as scattering amplitudes, for each object
individually.  These matrices incorporate the material characteristics
of each object individually, while the objects' relative positions and
orientations are described by universal translation matrices.  These
matrices connect the bases of wavefunctions, centered on each object,
in which the scattering amplitudes are calculated.  For the case of
two objects that are translationally invariant in the
$z$-direction, we  consider the energy per unit length
\begin{equation}
\frac{\cal E}{\hbar c L} =  \int_0^\infty \frac{d\kappa}{2 \pi}
\int_{-\infty}^\infty \frac{dk_z}{2 \pi}
\log \det \left({\mathbbm{1}} -
T_1  \, {\cal U}_{12} T_2  \,{\cal U}_{21}  \right)  \,,
\label{eqn:logdet}
\end{equation}
where the ${T}_{j}$ give the $T$-matrices for each object, while the
${\cal U}_{ij}=  {\cal U}_{ji}^\dagger$ give the translation matrices
from one object to another.  Here $k = i\kappa=\omega/c$ is the
magnitude of the wave vector for each possible fluctuation and $k_z$
is its $z$-component.  We can interpret this formula in terms of the
propagation of electromagnetic fluctuations between the objects:  The
$T$-matrices describe the reflection of fluctuating fields from a single
object, while the $U$-matrices propagate these fluctuations from one
object to another.  The determinant then combines all possible
reflections among the objects.  By evaluating the integral and
determinant numerically,  we can use Eq.~(\ref{eqn:logdet}) to find
exact results for the Casimir interaction energy.  However, this
expression also allows for a systematic approximation, in which we can
obtain simple analytic results.  Letting ${\cal N} = T_1
{\cal U}_{12} T_2 {\cal U}_{21}$, we can convert the
log-determinant to a trace-log, which we then expand as a Taylor
series,
\begin{equation}
\frac{\cal E}{\hbar c L} =
-\int_0^\infty \frac {d\kappa}{2 \pi}
\int_{-\infty}^\infty \frac {dk_z}{2 \pi}
\sum_{n=1}^\infty \frac{1}{n} \tr  {\cal N}^n \,.
\end{equation}
In this expansion, the $n^{\hbox{\tiny th}}$ term then gives the
contribution from $n$ reflections (back and forth{)}
 between the two objects.  These contributions typically fall
at least as fast as $n^{-(D+1)}$, where $D$ is the
dimension of space \cite{optical,wedge}.

We can realize the half-plane geometry as the limit of either a wedge
\cite{wedge} with zero opening angle or a parabolic cylinder
\cite{parabolic} with zero radius of curvature.  In the former
case the $T$-matrix is expressed in a basis indexed by
continuous imaginary angular momentum $\lambda$, while in the latter
case the $T$-matrix is given in a basis with
discrete channels $\nu = 0,1,2\ldots$.  Although we can compute either
the full determinant or the multiple reflection expansion using either
basis, the wedge basis is better suited to the reflection
expansion, because the associated translation matrix elements are
simpler and easier to handle analytically, while the parabolic
cylinder basis is better suited to the full determinant
calculation, because it is easier to calculate the determinant
when the matrix involves discrete rather than continuous indices.

\begin{figure}[htbp]
\hfill \includegraphics[width=0.3\linewidth]{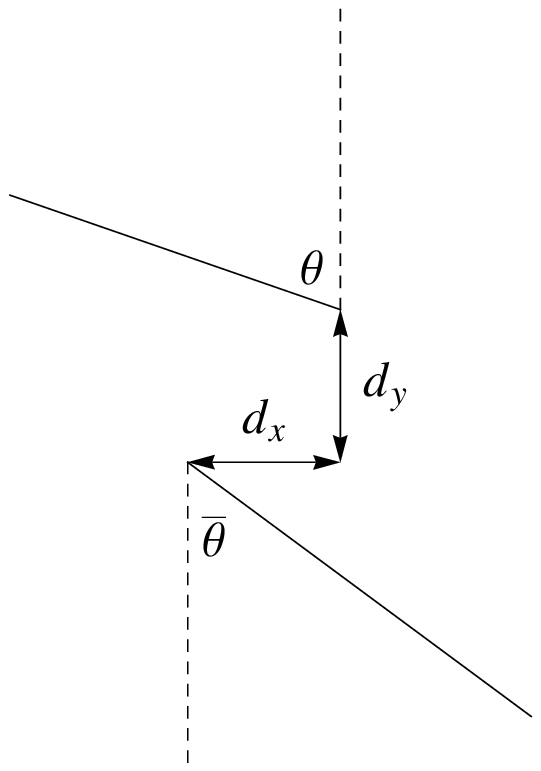}\hfill
\includegraphics[width=0.45\linewidth]{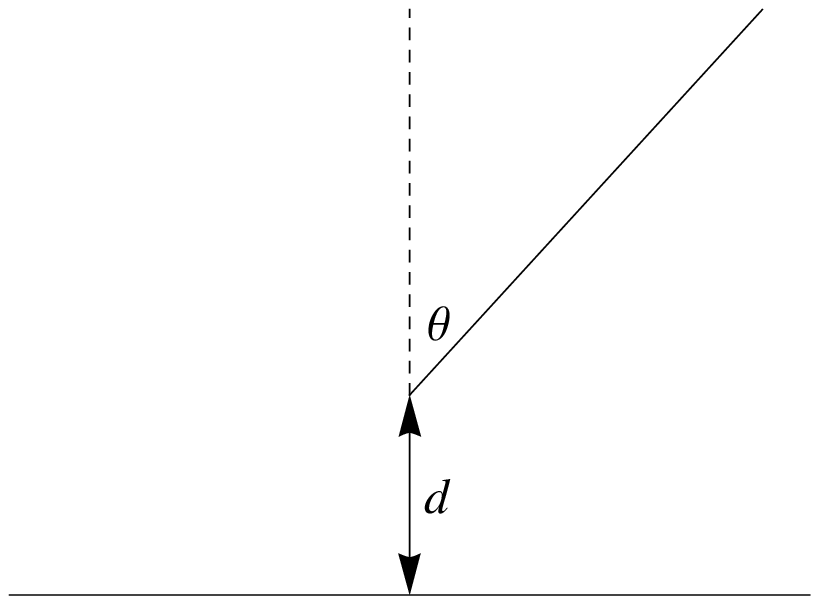} \hfill
\caption{
{Geometry for two half-planes (left panel) and
a half-plane opposite a plane (right panel).}
}
\label{fig:planegeom}
\end{figure}

\paragraph{Two Half-Planes}
We consider two half-planes, and restrict our attention to
configurations that are translation invariant in the $z$ direction.
We introduce a translation $d_y$ in the $y$ direction and $d_x$ in the
$x$ direction, and allow the upper and lower half-planes to rotate
around their edges away from the $y$-axis by angles $\theta$ and $\bar
\theta$ respectively, as shown in the left panel of
Fig.~\ref{fig:planegeom}.  This
description is redundant --- different parameter choices that lead to
the same physical configuration, a property we use to check our
calculations.   The numerical convergence of physically equivalent
configurations is not necessarily equivalent, however.  For example, in
the case of $\theta=\bar\theta = 0$, when both $d_x$ and $d_y$
increase, the Casimir interaction energy decreases, since the planes
are becoming further apart.  For the scattering bases we choose,
however, in  $d_y$ this effect appears directly through a decaying
exponential, while in $d_x$ it appears through the cancellation of an
oscillating integrand.  As a result, we need to maintain $d_y>0$,
while we can consider either sign of $d_x$.

For a wedge with zero opening angle, the $T$-matrix becomes
\cite{wedge}
\begin{equation}\label{eqn:TmatrixWedge}
{T}^{D/N}_\lambda = \mp \frac{1}{\cosh \lambda \pi} \,,
\end{equation}
where the channels are labeled by Dirichlet and Neumann boundary
conditions, corresponding to the two polarizations
of the electromagnetic waves, and by the
continuous index $\lambda$, the imaginary angular momentum.
We can then express the matrix ${\cal N}$ in the wedge basis as
\begin{equation}
\mathcal{N}{^{D/N}}_{\lambda \lambda'}(d_x,d_y,\theta,\bar\theta)
=\int_0^\infty d\lambda''  {T}_{{1}\, \lambda}^{D/N}
{\cal U}^{D/N}_{\lambda \lambda''}(d_x,d_y,\theta,\bar \theta)
{T}_{{2} \, \lambda''}^{D/N} {\cal U}_{\lambda''
\lambda'}^{D/N}(-d_x,d_y,-\theta,-\bar\theta)\, ,
\label{eqn:NmatrixWedge}
\end{equation}
where $\kappa$ and $k_z$ have been suppressed
because all the matrices involved are diagonal in these parameters.
Without loss of generality, $d_x$ can be set zero, in which case the
translation matrix $\mathcal U$ is given by
\begin{equation}
{\cal U}^{D}_{\lambda \lambda'}(0,d_y,\theta,\bar \theta)=
2\int_{-\infty}^\infty d k_x \frac{i}{k_y} \cosh(\lambda (\phi-\theta))
\cosh(\lambda' (\phi+\bar \theta))^*
\, e^{i k_y d_y} \,,
\label{eqn:UmatrixWedge}
\end{equation}
for the polarization corresponding to Dirichlet boundary
conditions, where $k_y=\sqrt{k^2-k_x^2-k_z^2} =
i\sqrt{\kappa^2+k_x^2+k_z^2}$ and $\phi=
\sin^{-1}\left({k_x}/{i\sqrt{\kappa^2+k_z^2}}\right)$. For
Neumann boundary conditions, the hyperbolic cosine functions
in Eq.~(\ref{eqn:UmatrixWedge}) are replaced by hyperbolic sine
functions.

In the parabolic cylinder basis, ${\cal N}$ becomes
\begin{equation}
{\cal N}_{\nu \nu'}(d_x,d_y,\theta,\bar\theta) =
\sum_{\nu''=0}^\infty {T}_{{1}\, \nu}
\,{\cal U}_{\nu \nu''}(d_x,d_y,\theta,\bar \theta)
\, {T}_{{2} \, \nu''}  {\cal U}_{\nu''
\nu'}(-d_x,d_y,-\theta,-\bar\theta) \,,
\end{equation}
where $T_{{\alpha}\, \nu} = -\sqrt{\frac{2}{\pi}} \nu!$ is the
$T$-matrix for the half-plane $\alpha =1,2$ in the scattering channel
$\nu=0,1,2,3\ldots$, with even $\nu$ corresponding to polarizations obeying
Dirichlet boundary conditions and odd $\nu$ corresponding to
polarizations obeying Neumann boundary conditions, and
the translation matrix is
\begin{equation}
{\cal U}_{\nu \nu'}(d_x,d_y,\theta,\bar \theta) =
\frac{1}{\sqrt{8 \pi \nu! \nu'!}}
\int_{-\infty}^\infty dk_x \frac{i}{k_y}
\frac{\left(\tan \frac{\phi + \theta}{2}\right)^{\nu}
\left(\tan \frac{\phi+\bar \theta}{2}\right)^{\nu'}}
{\cos \frac{\phi + \theta}{2} \cos \frac{\phi + \bar \theta}{2}}
e^{i k_x d_x}
e^{i k_y d_y} \,,
\end{equation}
where $\phi$ is defined as before.

We first evaluate the Casimir energy analytically to first order in the
reflection expansion.  Using Eqs.~(\ref{eqn:TmatrixWedge}),
(\ref{eqn:NmatrixWedge}), and (\ref{eqn:UmatrixWedge}), the
electromagnetic Casimir interaction energy becomes
\begin{equation}
\frac{{\cal E}}{\hbar c  L}= -\frac{1}{64\pi^3 d_y^2}
\left(\frac{8}{3} + 4 \csc \theta \csc \bar \theta
+ 4 (\theta \csc^2 \theta - \bar \theta \csc^2 \bar \theta) \csc
(\theta - \bar \theta) \right) + \cdots \,,
\label{eqn:parallel1}
\end{equation}
where without loss of generality we have taken $d_x=0$ and
the dots represent higher reflections.
Since the contribution at each reflection order is independent of the
scattering basis in which it is computed, this result can be obtained
using either the wedge or parabolic cylinder basis; however, the wedge
basis is more convenient because it yields integrals rather
than sums.  In the parabolic cylinder basis, one can obtain
the same results by writing the sum over $\nu$ as a geometric series
in $\tan \frac{\phi+\theta}{2}$.

\begin{figure}[htbp]
\includegraphics[width=0.5\linewidth]{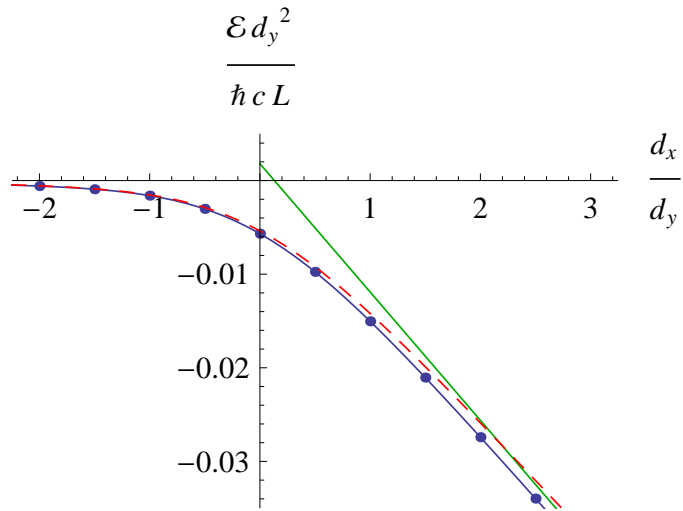}
\caption{Electromagnetic Casimir interaction energy per unit length
for overlapping planes as a function of horizontal displacement, in
units of the vertical separation $d_y$.  Circles are obtained from the
exact calculation in parabolic cylinder coordinates.  The solid line
connecting them is a rational function fit to guide the eye.  The
dashed line gives the analytic formula obtained by considering only
the first reflection, while the solid straight line gives the standard
parallel plane result for the overlap area, plus edge corrections.}
\label{fig:overlap}
\end{figure}

We are particularly interested in the case of parallel,
overlapping half-planes.  In this case, it is more convenient to fix
$\theta = \bar \theta = \pi/2$ and parameterize the configuration by
$d_x$ and $d_y$.  Then positive $d_x$ describes the width of the
overlap region, while negative $d_x$ describes a horizontal
displacement of the edges away from each other. The exact Casimir
interaction energy and the approximation
{to one reflection} for
this case are shown in Fig.~\ref{fig:overlap}. At large $d_x$, the
graph of $\frac{\mathcal E d_y^2}{\hbar c L}$ as a function of
$d_x/d_y$ becomes  asymptotic to a straight line with
slope $-\frac{\zeta(4)}{8 \pi^2}$,
consistent with the standard result for parallel planes:
The energy is linear in the exposed area and inversely
proportional to the cube of the separation distance,
\begin{equation}
\frac{\cal E}{\hbar c L} = -\frac{\pi^2 d_x}{720 d_y^3}
= -\frac{\zeta(4) d_x}{8 \pi^2 d_y^3}\,.
\label{eqn:Casimir}
\end{equation}
This result forms the basis for the proximity force
approximation (PFA), which sums over infinitesimal segments
treated as locally parallel planes \cite{PFA}.  The
$y$-intercept of the asymptote gives the correction --- beyond PFA ---
due to the interaction between each of the two edges and an infinite
plane.  It represents a constant shift
of the parallel plane result in the limit of large $d_x$, where the
edges are far from each other.  This correction is
positive --- it makes the overall Casimir
interaction energy less negative --- reflecting the suppression of
quantum fluctuations in the neighborhood of the sharp edge.
The dashed line in Fig.~\ref{fig:overlap} shows the result obtained
from the first reflection,
\begin{equation}\label{eqn:AnalyticForce}
\frac{{\cal E}}{\hbar c  L}= -\frac{1}{24\pi^3 d_y^2} \left[
\frac{1}{1+(d_x/d_y)^2} + 3 \left(1-i\frac{d_x}{d_y}
\log\frac{i-d_x/d_y}{\sqrt{1+(d_x/d_y)^2}}\right)
\right] + \cdots \,,
\end{equation}
which is Eq.~(\ref{eqn:parallel1}) specialized to the geometry of
overlapping parallel planes and expressed in terms of $d_x$ and
$d_y$.
Here the logarithm gives the arctangent in the appropriate
quadrant.  For large $d_x$, this result approaches the straight line
\begin{equation}\label{eqn:PFA1st}
\frac{\cal E}{\hbar c L} = -\frac{d_x}{8 \pi^2 d_y^3}\,.
\end{equation}
This equation represents the first reflection approximation to the
full result for parallel planes, which is obtained by replacing
$\zeta(4)$ in Eq.~(\ref{eqn:Casimir}) with the first term in the
expansion of the zeta function, \cite{optical}
\begin{equation}
\zeta(4) = \frac{\pi^4}{90} = \sum_{n=1}^\infty \frac{1}{n^4} =
1 + \frac{1}{16} + \frac{1}{81} + \cdots\,.
\end{equation}
We can see from Fig.~\ref{fig:overlap} that the first reflection
already gives an excellent approximation to the energy of the
overlapping planes.

Note that two half-planes also exert a lateral force on
each other (see Ref.~\cite{gedanken} for a gedanken experiment
based on this point). For positive $d_x$
(when the planes overlap), and to leading order, this force can be
obtained from the PFA, Eq.~(\ref{eqn:Casimir}), or, to one reflection,
from Eq.~(\ref{eqn:PFA1st}). The $y$-intercept, which quantifies the
interaction of each edge with an infinite plane, does not contribute
to the lateral force, since it is independent of the
distance between the two edges, $d_x$. However, the
interaction between the edges makes a contribution. The
analytic formula of Eq.~(\ref{eqn:AnalyticForce}) gives
\begin{equation}
\frac{F_x}{\hbar c L} = \frac{1}{8 \pi^2 d_y^3}-\frac{1}{6 \pi^2 d_x^3}
\left(1+\mathcal O(d_y^2/d_x^2)\right)+\cdots \hskip .5in d_x>0.
\end{equation}
The first term is simply PFA in the first reflection, while the
second term gives a negative correction, representing a suppression of
the force due to the interaction between the two edges. For
negative $d_x$, Eq.~(\ref{eqn:AnalyticForce}) gives
\begin{equation}
\frac{F_x}{\hbar c L} = -\frac{1}{6 \pi^2 d_x^3}
\left(1+\mathcal O(d_y^2/d_x^2)\right)+\cdots \hskip .5in d_x<0.
\end{equation}
We see that in the first reflection, the leading term of the lateral
force for $d_x<0$ is the same as the subleading term in the force for
$d_x>0$.  One can use an argument based on Babinet's principle to show
that this result holds for arbitrary $d_x/d_y$.  Applications of
Babinet's principle to the Casimir energy are studied extensively in
Ref.~\cite{Babinet}.

\paragraph{Half-Plane and Infinite Plane}
The edge correction can also be obtained by considering the limit in
which a half-plane becomes parallel to an infinite plane.  We
consider a half-plane separated by a distance $d$ from and tilted at
an angle $\theta$ from the perpendicular to an infinite plane, as
shown in the right panel of Fig.~\ref{fig:planegeom}.
As $\theta\to \pi/2$, the Casimir energy per
unit length diverges, since the energy is becoming proportional to the
area, but we can parameterize this divergence as
\begin{equation}
\frac{\cal E}{\hbar c L} = -\frac{c(\theta)}{\cos\theta} \frac{1}{d^2} \,,
\end{equation}
where $d$ gives the separation between the edge of the half-plane and
the infinite plane, and $\theta$ gives the angle between the half-plane
and the axis normal to the full plane.  Then for $\theta \to \pi/2$, where
the planes become parallel, we have
$c(\theta\to \pi/2)=c_\parallel/2+ \left(\theta - \pi/2\right)
c_{\hbox{\tiny edge}}$,
where $c_\parallel=\pi^2/720$ is the standard result for parallel
planes.  Then the slope $c_{\hbox{\tiny edge}}$ gives the correction
due to the interaction between the edge of the half-plane and the
infinite plane, in the limit where the half-plane is parallel to the
infinite plane.  Doubling this result gives the edge correction for
the case of overlapping planes, since there we have two edge-plane
interactions.

\begin{figure}[htbp]
\includegraphics[width=0.5\linewidth]{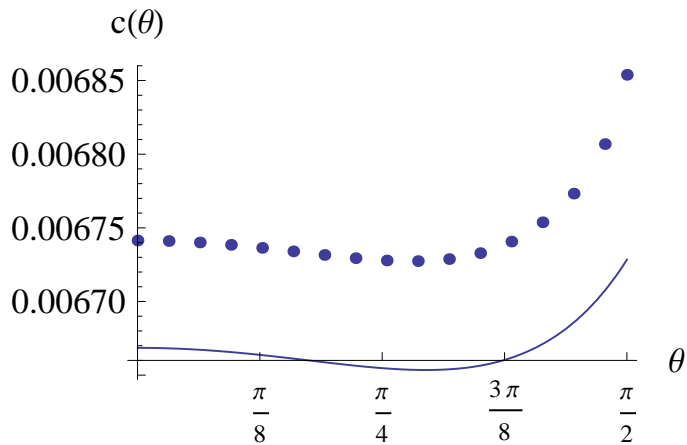}
\caption{The orientation dependence $c(\theta)$ for the
electromagnetic Casimir energy of a half-plane opposite a plane, as a
function of the tilt angle $\theta$.  Circles are obtained from the
exact calculation in parabolic cylinder coordinates, while the solid
line is obtained from the analytic expansion through
two reflections, Eq.~(\ref{eqn:edge2refl}).}
\label{fig:edge}
\end{figure}

The calculation for a half-plane opposite an infinite
plane proceeds analogously to the case of two planes, and has been
done in detail in Refs.~\cite{parabolic,wedge}.
In Fig.~\ref{fig:edge} we show the coefficient $c(\theta)$ using
both the exact numerical calculation in the parabolic cylinder basis
\cite{parabolic} and the approximation to two reflections in the wedge
basis \cite{wedge}, which is
\begin{equation}
\frac{{\cal E}}{\hbar c  L} = -\frac{1}{16\pi^2 d_y^2} \sec \theta
- \frac{1}{256\pi^3 d_y^2} \left(
\frac{4}{3}+ \csc^3\theta\sec\theta(2 \theta -
\sin 2 \theta) \right) + \cdots\,,
\label{eqn:edge2refl}
\end{equation}
where the dots indicate corrections from higher reflections (three
reflections or more).

The comparison of the exact numerical calculation in the parabolic
cylinder basis to the reflection expansion in the wedge basis shows
both the strengths and weaknesses of this approximation.
The expansion to two reflections for the energy of a
tilted half-plane opposite a plane captures the orientation dependence
$c(\theta)$ well.  The variation in this
quantity is small because of the near-cancellation of the
Dirichlet and Neumann contributions.  However, the relative error in the
edge correction, that is, the error in the derivative of $c(\theta)$
as $\theta\to \pi/2$, is large because while the neglected
contribution from the higher reflections is small, it varies
rapidly as the planes become parallel.  For $\theta = \pi/2$, the
reflection expansion to order $n$ gives the parallel-plane result with
$\zeta(4)$ truncated to the first $n$ terms in its series
representation, but this error decreases as the angle deviates from
$\pi/2$.  This behavior can be anticipated from a geometric optics
point of view, since as the planes become nonparallel, higher-order
reflections must propagate further as they reflect between the planes.
The analogous problem arises in extracting the edge correction from
the overlapping planes calculation, since again the asymptotic result
as $d_x\to\infty$ captures only the leading term in the zeta-function
series.

\paragraph{Thermal Corrections}
We can also apply these results to find the free energy in a system at
temperature $T$.  Then the integral over $\kappa$ is replaced by a sum
over Matsubara frequencies $\kappa_n = 2\pi n T/\hbar c$, where
$n=0,1,2,\ldots$ and the $n=0$ contribution is counted with a weight
of $1/2$.  Again, we keep only the first reflection.
For the case of a half-plane opposite a plane,
we obtain the free energy for each polarization
\begin{equation}
\frac{{\cal F}^{D/N}}{k_B T L} = -\frac{1}{2 \pi} {\sum_{
\kappa_n \ge 0}}' \int dk_x dk_z \ \frac{1}{4 \pi} e^{-2 d
\sqrt{\kappa_n^2+k_x^2+k_z^2}}\frac{1}{\sqrt{\kappa_n^2+k_x^2+k_z^2}}\left(\pm
\frac{\sqrt{\kappa_n^2+k_z^2}}{\sqrt{\kappa_n^2+k_x^2+k_z^2}} +
\frac{1}{\cos \theta}\right) + \cdots\,,
\label{eqn:perpT}
\end{equation}
where the dots represent higher reflections.  The first term in
parentheses cancels in the sum over polarizations, giving for
electromagnetism
\begin{equation}
\frac{{\cal F}{^{EM}}}{\hbar c L} =
-\frac{1}{32 \pi^2}\frac{1}{\cos \theta}
\frac{1}{\lambda_T d} \coth  \frac{d}{\lambda_T}
 =  -\frac{1}{32 \pi^2d^2}\frac{1}{\cos \theta}
\left[1+\frac{1}{3} \left(\frac{d}{\lambda_T}\right)^2 - \frac{1}{45}
\left(\frac{d}{\lambda_T}\right)^4+{\mathcal
O\left(\frac{{d}^6}{\lambda_T^6}\right)}\right]+ \cdots\,,
\end{equation}
where $\lambda_T = \hbar c /(2 \pi k_B T)$ is the thermal
wavelength.  The leading correction at small temperature is
proportional to $T^2$, but this term is independent of distance and
thus does not contribute to the force, for which the leading
correction starts at order $T^4$.  We note that in the case of a
scalar field with either Dirichlet or Neumann boundary conditions,
however, the first term in parentheses in Eq.~(\ref{eqn:perpT})
yields the contribution
\begin{equation}
\frac{\mathcal{F}{^{\pm}}}{\hbar c L} =\mp \frac{1}{16 \pi^3}
\frac{1}{\lambda_T d} \left[1+ \frac{d}{\lambda_T}\int_{1}^{\infty}
dt \, \frac{1}{t} \csch^2\left(\frac{d}{\lambda_T} t\right)
E(1-t^2)\right] \,,
\end{equation}
where $E(x)$ is the elliptic function.  This expression shows
nonanalytic behavior as $T\to 0$ proportional to $T^3 \ln T$, similar
to what was observed in Ref.~\cite{Gies}.

\paragraph{Acknowledgements}
We thank R.~Abravanel, T.~Emig, R.~L.~Jaffe, M.~Kardar,
M.~Kr{\"{u}}ger, S.~J.~Rahi, A.~Shpunt, and A.~Weber for helpful
conversations.  N.~G. thanks D.~Karabali and V.~P.~Nair for
discussions of thermal corrections.  We are especially grateful to
Professor Jaffe for a critical reading of the manuscript.
This work was supported in part by the U.\ S.\ Department of Energy
under cooperative research agreement \#DF-FC02-94ER40818 (MFM)
and by the National Science Foundation through grant PHY08-55426 (NG).


\begin{thebibliography}{9}

\bibitem{Casimir48}
H.~B.~G.~Casimir,
\newblock Proc.\ K.\ Ned.\ Akad.\ Wet.\ {\bf 51}, 793 (1948).

\bibitem{spheres}
T.~Emig, N.~Graham, R.~L.~Jaffe, and M.~Kardar,
\newblock Phys.\ Rev.\ Lett.\ {\bf 99}, 170403 (2007);
\newblock Phys.\ Rev.\ D {\bf 77}, 025005 (2008).

\bibitem{universal}
S.~J.~Rahi, T.~Emig, N.~Graham, R.~L.~Jaffe, and M.~Kardar,
\newblock Phys.\ Rev.\ D {\bf 80}, 085021 (2009).

\bibitem{parabolic}
N.~Graham, A.~Shpunt, T.~Emig, S.~J.~Rahi, R.~L.~Jaffe, and M.~Kardar,
\newblock Phys.\ Rev.\ D {\bf 81}, 061701(R) (2010).

\bibitem{wedge}
M.~F.~Maghrebi, S.~J.~Rahi, T.~Emig, N.~Graham, R.~L.~Jaffe, and
M.~Kardar, arXiv:1010.3223.

\bibitem{Maghrebi10-2}
M.~F.~Maghrebi, arXiv:1012.1060.

\bibitem{Khoshnoud}
F.~Khoshnoud, private communication.

\bibitem{Gies}
H.~Gies and K.~Klingmuller,
\newblock Phys.\ Rev.\ Lett.\  {\bf 97}, 220405 (2006);
A.~Weber and H.~Gies,
\newblock Phys.\ Rev.\ D {\bf 80}, 065033 (2009).

\bibitem{Kabat}
D.~Kabat, D.~Karabali, and V.~P.~Nair,
\newblock Phys.\ Rev. D {\bf 82}, 025014 (2010).

\bibitem{optical}
A.~Scardicchio and R.~Jaffe,
\newblock Nucl.\ Phys.\ B {\bf 704}, 552 (2005);
\newblock Nucl.\ Phys.\ B {\bf 743}, 249 (2006).

\bibitem{PFA}
B.~V.~Derjaguin, I.~I.~Abrikosova, and E.~M.~Lifshitz,
\newblock Q. Rev. Chem. Soc. {\bf 10}, 295 (1956).

\bibitem{gedanken}
G.~J. Maclay,
\newblock Phys. Rev. A {\bf 82}, 032106 (2010).

\bibitem{Babinet}
M.~F. Maghrebi, R.~Abravanel, and R.~L.~Jaffe,
\newblock to be published.

\end{thebibliography}
\end{document}